# SAM-dPCR: Real-Time and High-throughput Absolute Quantification of Biological Samples Using Zero-Shot Segment Anything Model

Yuanyuan Wei[1], Shanhang Luo[2], Changran Xu[3], Yingqi Fu[1], Yi Zhang[4], Fuyang Qu[1], Guangyao Cheng[1], Yi-Ping Ho[1, 5, 6, 7], Ho-Pui Ho[1, *], Wu Yuan[1, *]

[1] Department of Biomedical Engineering, The Chinese University of Hong Kong, Shatin, Hong Kong SAR, 999077, China. E-mail: wyuan@cuhk.edu.hk; aaron.ho@cuhk.edu.hk

[2] Department of Biomedical Engineering, National University of Singapore, 119077, Singapore

[3] Department of Computer Science and Engineering, The Chinese University of Hong Kong, Shatin, Hong Kong SAR, 999077, China

[4] Department of Electronic Engineering, The Chinese University of Hong Kong, Shatin, Hong Kong SAR, 999077, China

[5] Centre for Biomaterials, The Chinese University of Hong Kong, Hong Kong SAR, 999077, China

[6] Hong Kong Branch of CAS Center for Excellence in Animal Evolution and Genetics, Hong Kong SAR, 999077, China

[7] State Key Laboratory of Marine Pollution, City University of Hong Kong, Hong Kong SAR, 999077, China

[*]Correspondence: wyuan@cuhk.edu.hk; aaron.ho@cuhk.edu.hk.

**Abstract**

Digital PCR (dPCR) has revolutionized nucleic acid diagnostics by enabling absolute quantification of rare mutations and target sequences. However, current detection methodologies face challenges, as flow cytometers are costly and complex, while fluorescence imaging methods, relying on software or manual counting, are time-consuming and prone to errors. To address these limitations, we present SAM-dPCR, a novel self-supervised learning-based pipeline that enables real-time and high-throughput absolute quantification of biological samples. Leveraging the zero-shot SAM model, SAM-dPCR efficiently analyzes diverse microreactors with over 97.7% accuracy within a rapid processing time of 3.16 seconds. By utilizing commonly available lab fluorescence microscopes, SAM-dPCR facilitates the quantification of sample concentrations ranging from 0.74 to 17.49× $10^3$ copies $\mu L^{-1}$ for target nucleic acid templates. The accuracy of SAM-dPCR is validated by the strong linear relationship ($R^2$ = 0.9939) observed between known and inferred sample concentrations. Additionally, SAM-dPCR demonstrates versatility through comprehensive verification using various samples and reactor morphologies. This accessible, cost-effective tool transcends the limitations of traditional detection methods or fully supervised AI models, marking the first application of SAM in nucleic acid detection or molecular diagnostics. By eliminating the need for annotated training data, SAM-dPCR holds great application potential for nucleic acid quantification in resource-limited settings.

## 1. Introduction

The ongoing COVID-19 pandemic has highlighted the urgent need for accurate and rapid quantification of biological samples, particularly for diagnostic purposes[1–3]. Absolute quantification, which provides precise numerical expression levels, is crucial for improving the sensitivity and effectiveness of detecting rare templates, such as viral particles or low-abundance biomarkers[4]. Accurate measurement of biological samples is essential for various applications in genomics[5,6], proteomics[7,8], and molecular diagnostics[9,10].

Digital PCR (dPCR) is an advanced technology that enables the detection and quantification of rare mutations and target sequences[11]. Unlike conventional PCR or quantitative PCR (qPCR), which rely on relative quantification, dPCR allows for highly sensitive absolute quantification of nucleic acids without the need for standard curves[4,12]. eliminates amplification bias and enhances reproducibility. However, current analysis methodologies for dPCR face significant challenges, including high cost, complexity, time consumption, and potential errors[13–15].

In dPCR, the sample undergoes dispersion into tens of thousands of compartments for amplification[4,16,17]. The inferred concentration is ascertained by employing Poisson statistics to analyze the proportions of positive (fluorescent) and negative (little to no fluorescence) droplets[18,19]. Consequently, the precise identification of positive droplets in dPCR images is crucial for preserving the quantitative analysis accuracy of target nucleic acids. Flow cytometers, frequently regarded as the gold standard for quantification, are costly and necessitate specialized training for operation and analysis[20]. Conversely, fluorescence imaging, which depends on software or manual counting, is time-intensive and susceptible to human errors[21].

Despite the remarkable strides made through deep learning in bio-analysis[22–27], its application in digital Polymerase Chain Reaction (dPCR) analysis remains fraught with challenges. Traditional supervised machine learning algorithms, such as Mask R-CNN[28] and Yolov5[29,30], have been successful in automating droplet reading and analysis. However, these methods are hampered by several limitations. Firstly, these supervised machine learning algorithms require clean or 'ground truth' data for training, which can only be satisfied through labor-intensive data collection and manual annotation[31,32]. This process often leads to a marked gap between training and inferring domains due to the inherent variability in dPCR data. Moreover, the precondition of interframe continuity in these supervised methods may limit their ability to accurately visualize rapid transformations, such as changes in droplet diameters or primer sets[1,33]. This limitation necessitates further model training to accommodate these new experimental settings, thereby impeding the comparability and reproducibility of results across different laboratories and platforms.

To address these limitations, we introduce the SAM-dPCR, a novel and high-throughput algorithm for real-time absolute quantification of biological samples. SAM-dPCR combines the cutting-edge Zero-Shot Segment Anything Model (SAM) open-sourced by Meta[34] with the efficiency and accuracy of dPCR technology, offering a groundbreaking and generative solution. By leveraging a large-scale dataset and employing a deep neural network architecture, SAM stands out for its exceptional generalization capabilities, seamlessly adapting to diverse imaging scenarios without extensive retraining or fine-tuning. SAM-dPCR eliminates the need for labor-intensive data annotation, thereby streamlining the quantification process and significantly reducing analysis time. Furthermore, SAM's self-supervised nature enables SAM-dPCR to adapt to different appearances and characteristics of objects of interest across variations in digital PCR image quality, accommodating varying experimental conditions and sample types. SAM-dPCR not only enhances the analysis and interpretation of digital PCR results but also improves the reliability and utility of this vital molecular biology technique.

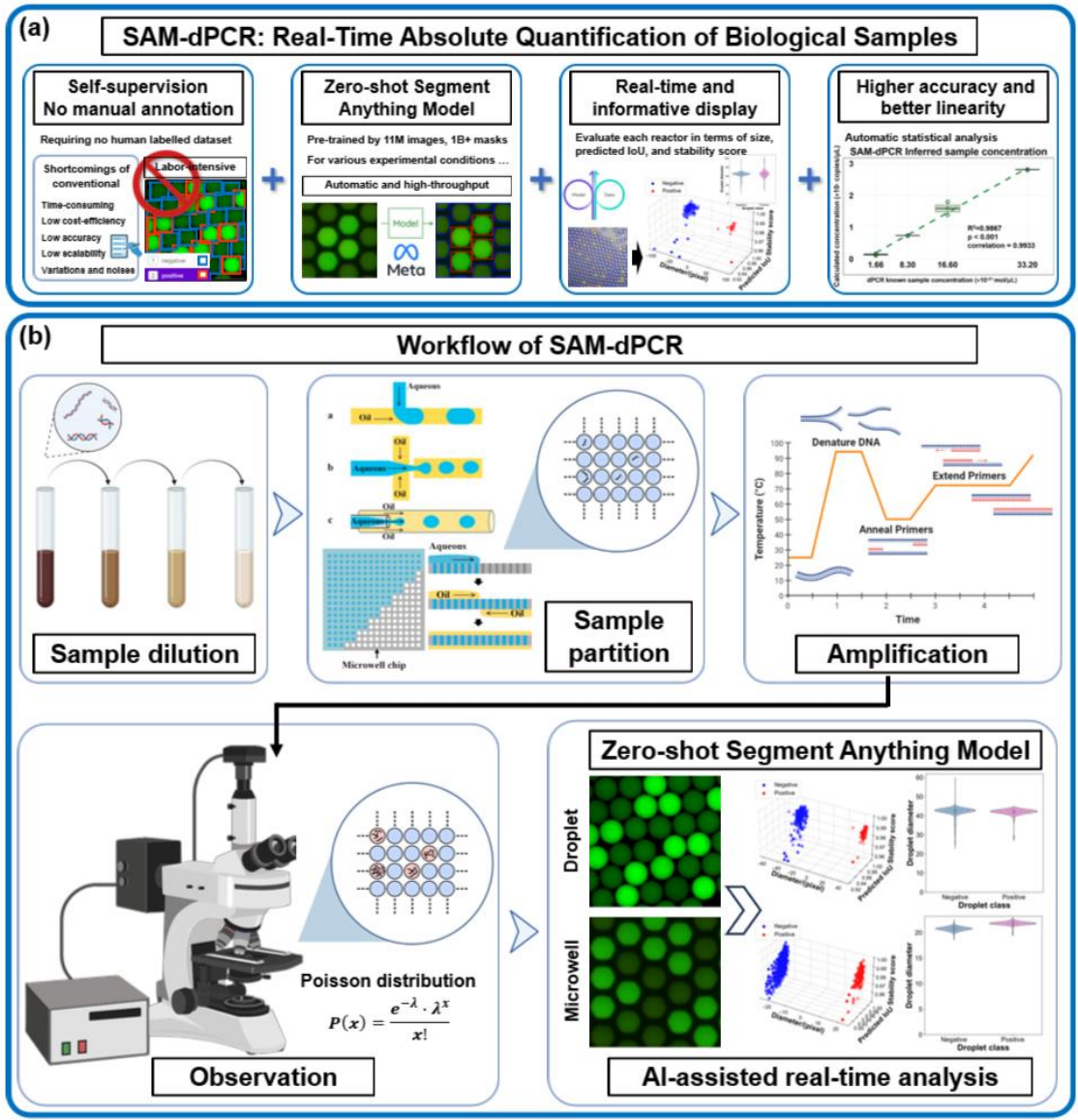

**Fig. 1 Illustration and workflow of SAM-dPCR for real-time digital nucleic acid assays.** The process begins with sample super dilution, followed by partitioning the PCR mixture into 20,000+ microreactors. Post-PCR amplification, ~2,000 microreactors are imaged and analyzed by the SAM-dPCR algorithm. The algorithm automatically segments microreactors and evaluates each mask in terms of size, predicted IoU, and stability score. Microreactors are classified as positive

(red) or negative (blue) based on fluorescence intensity. The template concentration is estimated by performing a statistical analysis that involves fitting the proportion of positive reactors to the Poisson distribution. The successful application of both droplet and microwell digital PCR is depicted.

2. **Results**

The SAM-dPCR algorithm was developed to enable automatic and high-throughput absolute quantification of target nucleic acid templates, eliminating the need for annotated training data. The samples were subjected to serial dilution and partitioning using droplet-based and microwell-based platforms respectively, as illustrated in **Fig. 1**. Utilizing a customized lab-on-a-chip system, each sample generated over 20,000 monodispersed droplets, with individual droplets measuring 46.37 ± 1.64 µm (equivalent to a volume of 52.20 pL). Following a three-temperature droplet digital PCR (ddPCR) amplification process, approximately 2,000 microreactors were imaged using FITC fluorescence microscopy. Real-time processing of the captured images was performed through a graphical user interface (GUI) and analyzed using the SAM-dPCR algorithm. The zero-shot SAM model, in combination with a fluorescence intensity-based classification algorithm, facilitated automatic image segmentation and classification. Droplets containing templates were classified as positive (red), while those without templates were classified as negative (blue), based on sequential frame analysis at a rate of one frame per second. Each segmented mask was evaluated simultaneously based on its diameter, predicted Intersection over Union (IoU), and stability score. Subsequently, the template concentration was estimated by performing a statistical analysis that involves fitting the proportion of positive reactors to the Poisson distribution.

The SAM-dPCR algorithm was validated across varied sample concentrations to analyze its performance and effectiveness. **Fig. 2(a)** and **(b)** illustrate the application of SAM-dPCR to both droplet and microwell dPCR experiments, respectively. The figures display overlaid segmentation masks on sample images, which are automatically annotated by the zero-shot SAM model. These masks delineate the microreactors, enabling further classification and analysis. The analysis results yielded inferred concentrations ranging from 0.74 to 17.49× $10^3$ copies µL$^{-1}$, demonstrating a strong linearity correlation ($R^2$ = 0.9939) with known sample concentrations. This robust correlation underscores the precision and validity of our SAM-dPCR approach. **Fig. 2(b)** presents similar results for microwell dPCR experiments, analyzed using SAM-dPCR algorithm without necessitating transfer learning, modifications, or retraining. Again, a strong linearity correlation ($R^2$ = 0.9867) was observed, with concentrations inferred ranging from 0.14 to 2.81 × $10^3$ copies µL$^{-1}$, reinforcing the efficacy of our approach. To validate the approach, here our templates utilized the Seahorse (*Hippocampus kuda*) genome, specifically the cytochrome c oxidase subunit I (*COI*) region with an amplicon size of 206 bp, which was serially diluted to concentrations of 0.4 pg, 4 pg, and 40 pg per 20 µL PCR system. The microwell dPCR dataset was prepared similarly, with concentrations ranging from 1.66 × $10^{-15}$ mol/L to 1.66 × $10^{-13}$ mol/L and a reaction-well volume of 755 pL. The amplified templates included two types of double-stranded genes, blaNDM and blaVIM, which control the expression of β-lactamases (bla), an antibiotic agent for carbapenems. Further details of the dPCR experiments can be found in the **Materials and Methods** section.

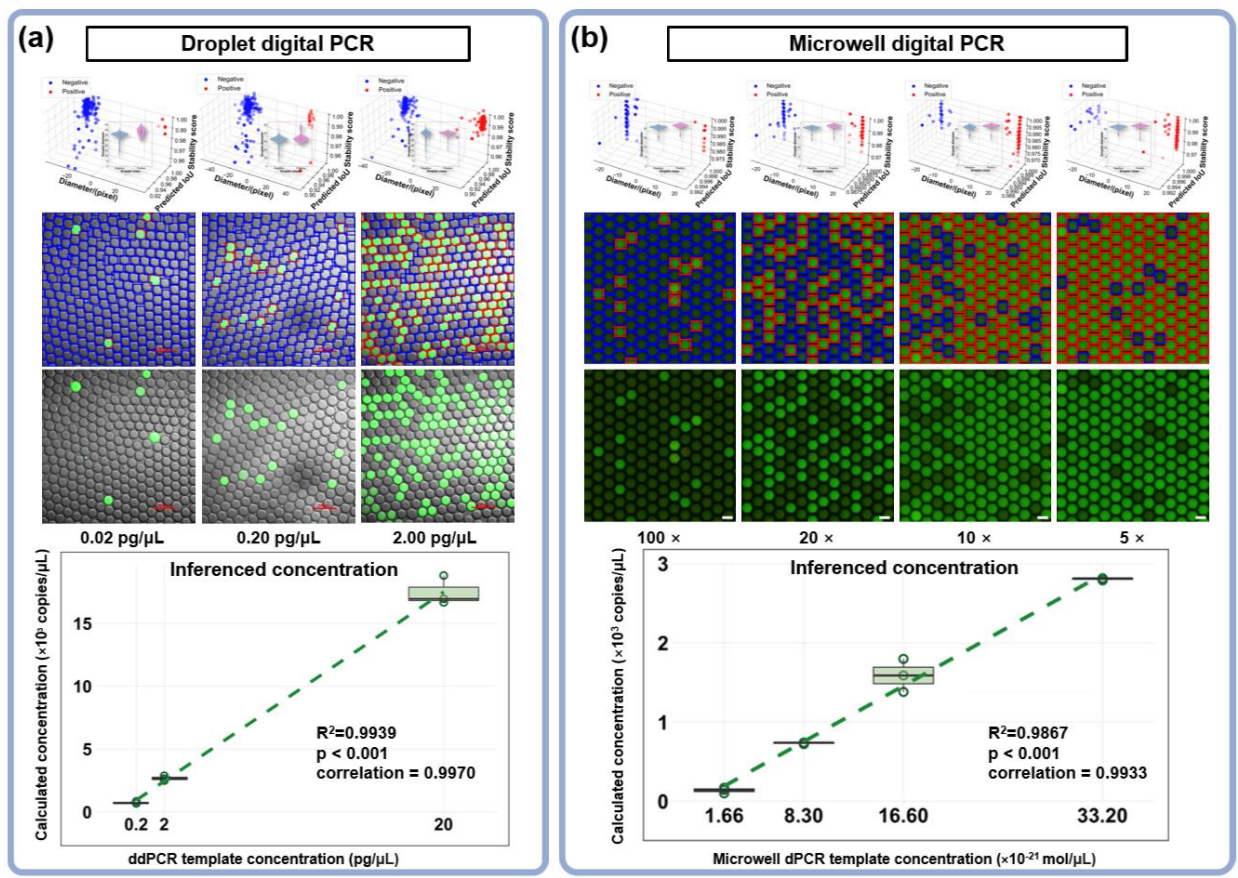

**Fig. 2 Efficacy of the SAM-dPCR algorithm across dPCR tasks with varying sample concentrations.** The figure displays sample images overlaid with masks from benchtop dPCR experiments, including both droplet PCR (a) and microwell dPCR (b) with serially diluted templates. The segmentation masks, automatically annotated by the zero-shot SAM model, outline the microreactors for further classification and analysis. Sample concentrations are determined by fitting the analysis results to a Poisson distribution. The strong linearity correlation, demonstrated by the linear regression equations, confirms the validity and precision of the SAM-dPCR algorithm. (Scale bar = 200 μm).

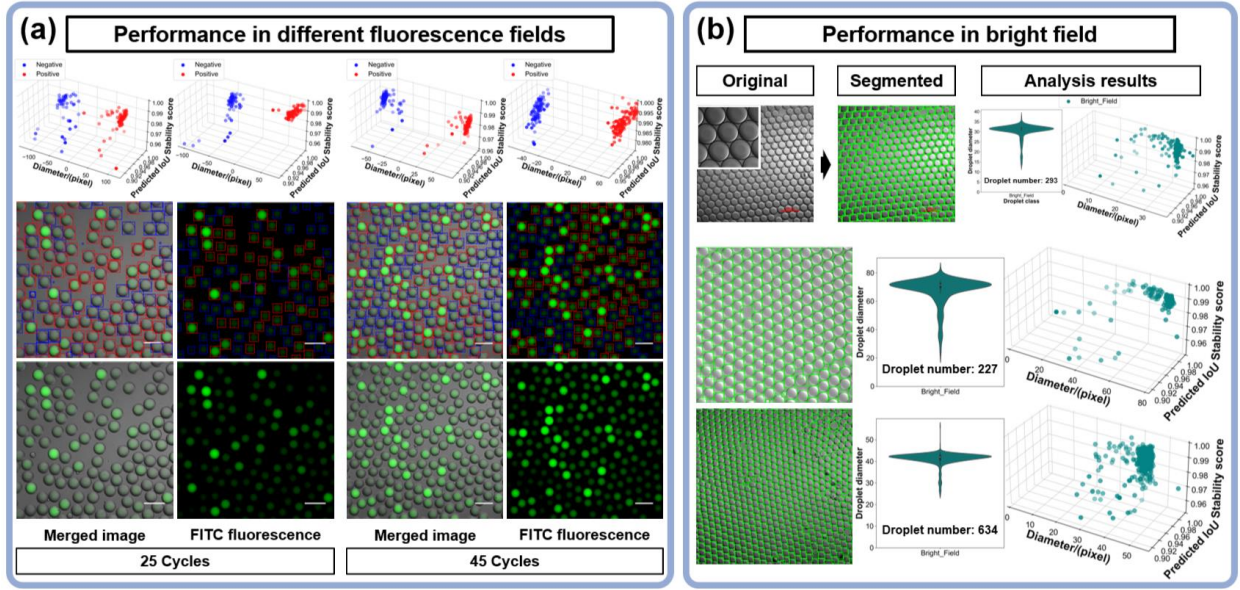

**Fig. 3 Efficacy of the SAM-dPCR algorithm under varied imaging and thermal cycling conditions.** Sample images from benchtop ddPCR experiments, including FITC fluorescence images, bright field images, and merged images, are overlaid with automatically annotated masks. (a) The post-45 cycle results generally exhibit a higher number of classified positive droplets compared to 25 cycles. FITC images show reduced size variation and a lower FAR, while merged images display a lower FRR. (Scale bar = 100 μm). (b) Automatic segmentation and analysis of bright field images using the zero-shot SAM model. The analysis of droplet diameter from bright field images demonstrates higher accuracy and uniformity compared to fluorescence field images, which can be attributed to the elimination of fluorescence scattering.

The adaptability of the SAM-dPCR algorithm was assessed under diverse imaging and thermal cycling conditions. The results, depicted in **Fig. 3**, demonstrate the algorithm's effectiveness when applied to fluorescence, bright, and merged field dPCR images. Analysis of fluorescence and merged images using SAM-dPCR (**Fig. 3(a)**) revealed that FITC fluorescence images exhibited less size variation and a lower false accept rate (FAR), indicating a higher level of masks IoU and stability score. Conversely, merged images displayed a lower false rejection rate (FRR), indicating improved droplet classification results. Furthermore, the algorithm's performance was evaluated through testing on images with varied fluorescence intensities, which were obtained under different PCR amplification conditions. Notably, PCR cycling for 25 and 45 cycles showed that the latter condition led to a greater number of droplets being identified and classified as positive, highlighting the adaptability of SAM-dPCR to varying thermal cycling conditions. In **Fig. 3(b)**, SAM-dPCR was successfully applied to bright field images without the need for modifications or retraining. The algorithm accurately segmented droplets ranging from dozens to hundreds in number. Moreover, the analysis of droplet diameter from bright field images demonstrated higher accuracy compared to fluorescence field images, which can be attributed to the elimination of fluorescence scattering. These results underscore the robustness and generalization capability of SAM-dPCR under varied experimental conditions.

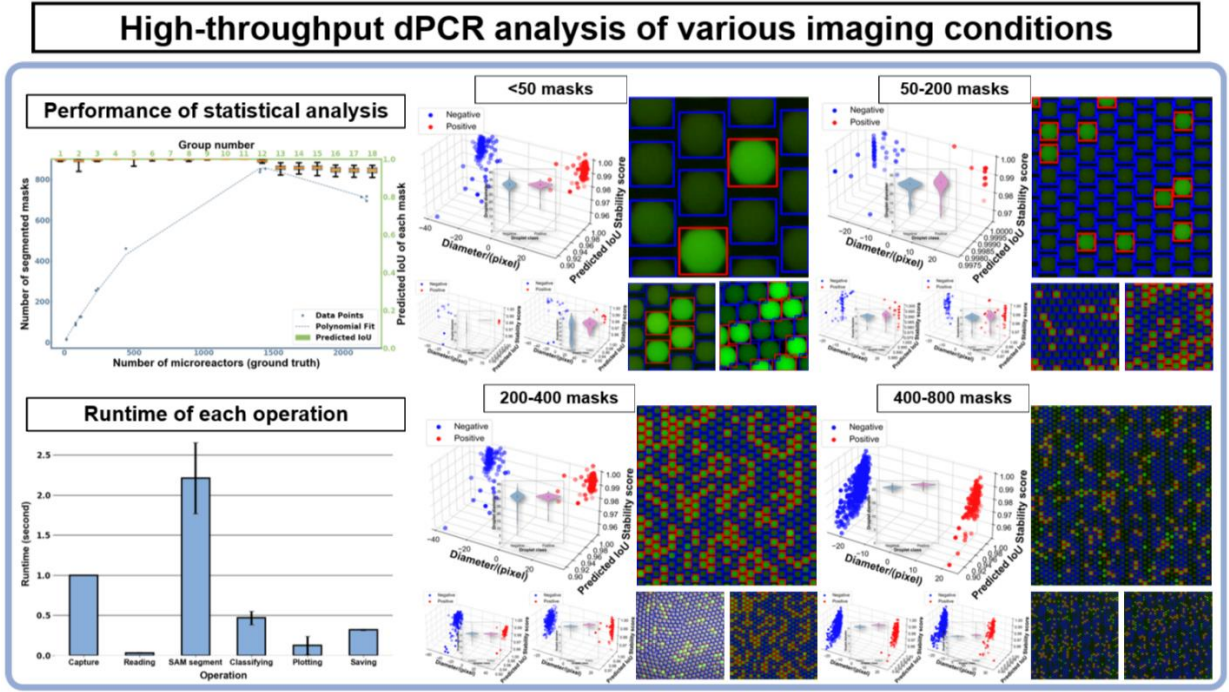

**Fig. 4 Evaluation of the SAM-dPCR algorithm's robustness and versatility through dPCR image analysis testing.** (a) and (b) Sample images overlaid with masks, derived from our newly curated dataset, featuring both droplet and microwell dPCR benchtop experiments. Images categorized based on the number of masks per image, ranging from fewer than 50 to 400-800. SAM-dPCR demonstrates robust performance when handling between 200 to 600 reactors per image, with a strong correlation between segmented mask count and ground truth. Predicted Intersection over Union (IoU) consistently exceeds 93% within this range, achieved without manual annotation or a training dataset. SAM-dPCR generates comprehensive three-dimensional plots, enabling simultaneous visualization of hundreds of reactors within 3.16 seconds. Manual counting measurement served as the ground truth for evaluation. (c) The breakdown of SAM-dPCR's runtime. The algorithm integrates seamlessly into common fluorescence microscopes, offering automatic and efficient analysis, and reducing time and cost requirements significantly.

The robustness and versatility of the SAM-dPCR algorithm were evaluated through a series of dPCR image analysis experiments, as depicted in **Fig. 4**. The dataset comprises a collection of 1024×1024 fluorescence images obtained from a diverse range of dPCR experiments, representing various sample types, target genes, and experimental conditions. This diversity allows for the robust testing and evaluation of our algorithm's performance. The images were categorized based on the number of masks per image, ranging from fewer than 50 to 400-800, as shown in **Fig. 4(a)** and **(b)**. Here manual counting measurement served as the ground truth for evaluation. Our findings demonstrate that the SAM-dPCR algorithm exhibits robust performance when handling between 200 to 600 reactors per image. We observed a strong correlation between the segmented mask count and the ground truth, with the predicted IoU of masks consistently exceeding 93% within this range. Notably, this performance was achieved without the need for manual annotation or a training dataset, owing to SAM's pre-training on over 1 billion masks from 11 million licensed and privacy-respecting images[34]. Additionally, the SAM-dPCR algorithm generates comprehensive three-dimensional plots for each image, enabling visualization of hundreds of reactors simultaneously within a remarkable time frame of 3.16 seconds. In contrast, manual counting takes over 200 seconds to count approximately 300 droplets. Moreover, conventional methods, such as manual counting or software ImageJ do not provide this level of detail, as they solely offer user-calculated sample concentration. A detailed breakdown of our algorithm's runtime is provided in **Fig. 4(c)**. The SAM-dPCR analysis process encompasses four main computational operations: (1) reading images from the API or SDK cable of the fluorescence microscope, taking approximately 0.03 seconds per image; (2) segmenting and classifying microreactors using SAM-dPCR, requiring less than 2.21 seconds and 0.47 seconds per image, respectively; (3) plotting and displaying the results, taking approximately 0.13 seconds per image; and (4) outputting the labeled images and analyzed results into a designated folder, necessitating around 0.04 seconds per image. Regarding adaptability, SAM-dPCR algorithm can seamlessly integrate into common laboratory fluorescence microscopes. However, commercialized software often necessitates specific facilities. For instance, Bio-Rad's QX Manager is limited to Bio-Rad Droplet Digital PCR Systems, including QX600, QX600 AutoDG, QX200, and QX200 AutoDG. In contrast, our SAM-dPCR does not require specific facilities or trained operators. Our approach demonstrates automatic and efficient analysis capabilities, significantly reducing the time and cost required.

We evaluated the performance of our algorithm, SAM-dPCR, by comparing it with the fully supervised Deep-qGFP model, as illustrated in **Fig. 5**. Deep-qGFP was trained using a Yolo-V5m model and the Region Proposal Network (RPN) on over 200 manually labeled ddPCR datasets. To establish a benchmark, we employed manual counting and ImageJ measurements as the ground truth. Our results indicate that SAM-dPCR outperforms Deep-qGFP in terms of droplet diameter measurement and dPCR image analysis. In **Fig. 5(a)** and **Table 1**, SAM-dPCR outperformed Deep-qGFP by accurately estimating droplet diameter with an average deviation of only 5.177 pixels, while Deep-qGFP consistently underestimated droplet diameter with an average deviation of 24.893 pixels. This discrepancy can be attributed to the difference between the training dataset and the testing dataset, which contains droplets with non-uniform fluorescence distribution due to varied experimental conditions. SAM-dPCR also achieved a higher accuracy of 97.658% in droplet number counting compared to Deep-qGFP's 96.23% as **Fig. 5(b)** depicts. SAM-dPCR also exhibits a lower positive droplet count error of 1.548% (equivalent to 3.5 droplets), while Deep-qGFP has an error of 9.491% (equivalent to 13.75 droplets). **Fig. S1 (Supplementary Information**) further supports these results across different testing datasets. These findings highlight the limitations of fully supervised learning approaches that rely on data-driven priors learned from paired noisy and clean measurements. In contrast, SAM-dPCR provides higher accuracy and adaptability, underscoring its superiority over conventional algorithms.

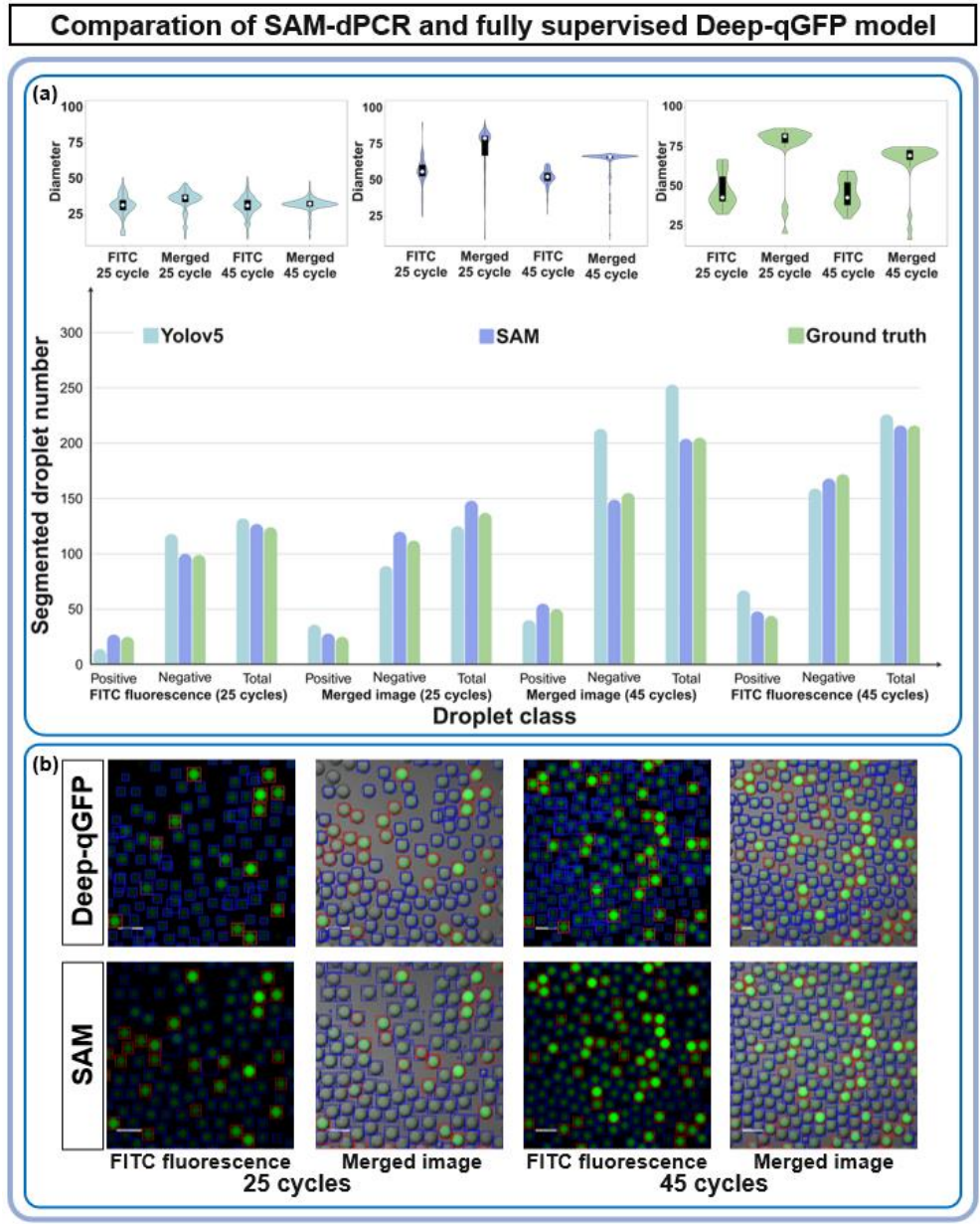

**Fig. 5 Performance evaluation of SAM-dPCR compared to the fully supervised Deep-qGFP model.** (a) SAM-dPCR accurately estimated droplet diameter with an average deviation of 5.177 pixels, while Deep-qGFP consistently underestimated droplet diameter with an average deviation of 24.893 pixels. This discrepancy can be attributed to non-uniform fluorescence distribution in the testing dataset due to varied experimental conditions. (b) SAM-dPCR achieved a higher accuracy of 97.658% in droplet number counting compared to Deep-qGFP's 96.23%. SAM-dPCR also exhibited a lower positive droplet count error of 1.548% (equivalent to 3.5 droplets), while Deep-qGFP had an error of 9.491% (equivalent to 13.75 droplets). Notably, SAM-dPCR competes with or even surpasses the fully supervised Deep-qGFP results, despite not requiring a manually annotated training dataset. Manual counting and ImageJ measurement served as the ground truth for comparison. (Scale bar = 100 μm).

**Table 1. Comparative analysis of our algorithm and fully supervised Deep-qGFP model.**

| Metric | Our algorithm | Deep-qGFP |
|---|---|---|
| Require annotated dataset | **No** | Yes |
| Droplet diameter error | **5.177 pixels** | 24.893 pixels |
| dPCR analysis accuracy | **97.658%** | 96.230% |
| Positive droplet count error | **1.548%** | 9.491% |

Our SAM-dPCR algorithm exhibits robust generalization capabilities when directly applied to a range of microreactor-based biological applications, as illustrated in **Fig. 6**. These microreactors, including droplets, microwells, and agaroses, are enhanced with DNA intercalating dyes, such as SYBR Green or EvaGreen, for visualization. Our SAM-dPCR analysis algorithm has been successfully employed in droplet single-cell sequencing (**Fig. 6 (a)**), agarose digital PCR (**Fig. 6 (b)**), and droplet-based digital quantification of bacterial suspensions (**Fig. 6 (c)**). In each of these scenarios, the microreactors are automatically and accurately segmented and classified, with the resulting data plotted and concentrations calculated accordingly. Notably, our approach offers significant advantages in terms of efficiency and accessibility compared to conventional manual counting methods. Furthermore, the results highlight the broad generalization capability of the SAM-dPCR algorithm, surpassing reported fully supervised AI models. To the best of our knowledge, this represents the first successful implementation of a self-supervised AI model in various droplet microfluidics scenarios, including ddPCR, droplet single-cell sequencing, and droplet-based bacterial quantification.

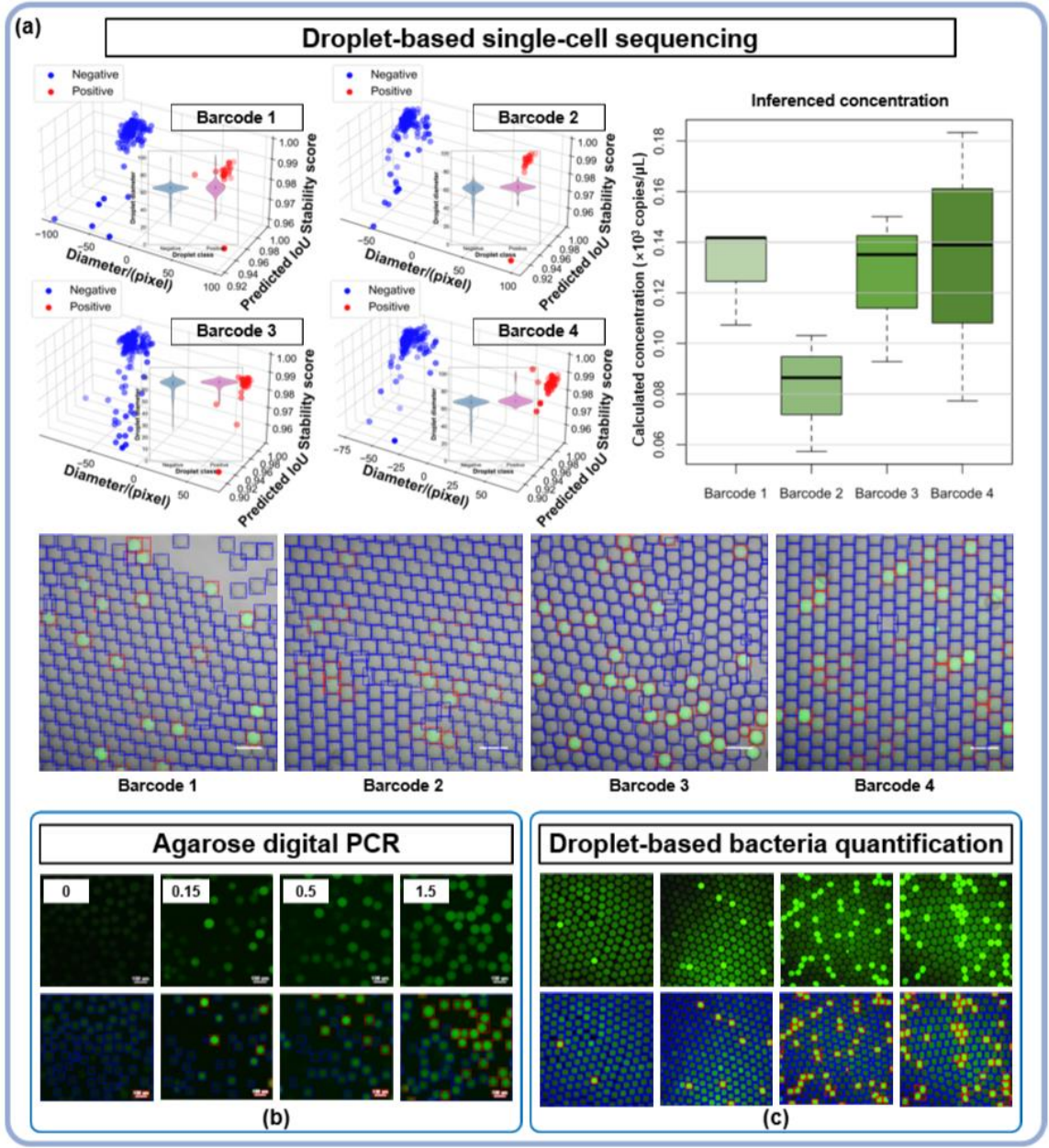

**Fig. 6 Versatile applicability of SAM-dPCR demonstrated across various biological applications.** The SAM-dPCR algorithm was successfully applied in diverse contexts, supplemented with DNA intercalating dyes for visualization such as SYBR Green or EvaGreen. (a) Droplet detection and classification results for droplet single-cell sequencing with varied barcoding. Statistical validation of results was performed, analyzing over 2,000 droplets per experimental condition, with inferred concentrations plotted. (Scale bar = 100 μm). (b) Image analysis results showcasing representative frames from agarose-based digital PCR with different encapsulation rates. (Scale bar = 100 μm). (c) Image analysis results of representative frames for droplet-based digital bacterial quantification under varying suspension conditions. (Scale bar = 100 μm).

## 3. Discussion and Conclusion

Digital PCR (dPCR) has revolutionized molecular biology by providing unparalleled precision, sensitivity, and absolute quantification of nucleic acids. However, traditional analysis methods such as manual counting or flow cytometry face challenges including high costs, complexity, and limited accessibility. Although AI-assisted automatic analysis techniques have been reported, existing models are fully supervised and limited to specific training datasets, making it difficult for rapid transformations, such as changes in droplet diameters or primer sets. Consequently, the comparability and reproducibility of results across different laboratories and platforms are compromised.

To address these limitations, we introduce SAM-dPCR, which integrates the Zero-Shot Segment Anything Model (SAM) with a fluorescence intensity-based classification algorithm. SAM-dPCR efficiently analyzes diverse dPCR images with over 97.7% accuracy within a rapid processing time of 3.16 seconds. It can be seamlessly integrated into commonly available lab fluorescence microscopes, enabling the quantification of sample concentrations ranging from 0.74 to 17.49× $10^3$ copies $\mu L^{-1}$ for target nucleic acid templates. The accuracy of SAM-dPCR is validated by the strong linear relationship ($R^2$ = 0.9939) observed between known and inferred sample concentrations.

In this study, we demonstrate the remarkable benefits and versatility of SAM-dPCR in absolute quantification by applying it to various DNA intercalating dye-labeling scenarios, such as SYBR Green and EvaGreen. These scenarios encompass droplet-based, microwell-based, and agarose-based applications commonly employed in a wide range of biological experiments. Our image datasets span a diverse range of droplet and microwell patterns, sample types, target genes, and experimental settings, highlighting the versatility of SAM-dPCR across various dPCR applications and experimental conditions. SAM-dPCR operates in two modes: offline and real-time. In offline mode, images and videos are collected and stored for post-experiment analysis, allowing for parameter fine-tuning. In real-time mode, the captured image and analysis results are displayed on our GUI simultaneously.

The performance of SAM-dPCR is influenced by various factors, including the model architecture and image quality parameters such as focusing, luminance, and resolution. Here we acquired SAM due to its remarkable generalization capabilities, enabling it to adapt seamlessly to diverse imaging scenarios without extensive retraining or fine-tuning. The SAM model can be loaded with three different encoders: ViT-B, ViT-L, and ViT-H. While ViT-H exhibits substantial improvements over ViT-B, the gains over ViT-L are marginal. These encoders have different parameter counts, with ViT-B having 91M, ViT-L having 308M, and ViT-H having 636M parameters. This difference in size also affects the speed of data processing. To maximize accessibility and simplify the model architecture, we acquired the ViT-B encoder to develop the SAM-dPCR algorithm. Additionally, we optimized imaging settings, including focusing, exposure conditions, magnifications (ranging from 4× to 20×), droplet sizes (diameter ranging from 18 to 114 pixels), and image resolutions (ranging from 256×256 to 4140×3096 pixels), for both bright fields and fluorescence fields. Our experiments demonstrated that fluorescence noise caused by fouling or specks of dust did not interfere with the segmentation results, as it was much smaller than the microreactors in size. We also found minimal spatial variation in the background signal and the exposure allowed for a certain level of light transmission loss. The SAM-dPCR pipeline utilizes disposable microfluidic chips for routine background signal evaluations.

We determined that the SAM-dPCR algorithm exhibits robust performance when handling between 200 to 600 reactors per image, as shown in **Fig. 4**. However, reactor size measurement error occurs heavily when the number of masks is smaller than 100 (**Fig. S2(a), Supplementary Information**) due to incomplete microreactors in the testing image. Additionally, miss-detection occurs heavily when the number of masks is larger than 600 (**Fig. S2(b), Supplementary Information**). This can be attributed to the limitation of the training dataset of SAM, in which there are approximately 100 masks per image on average. This issue can be addressed by fine-tuning the SAM model on more specific downstream dPCR image segmentation tasks or cropping the image into smaller regions with microreactor numbers less than 400 of the testing image.

In conclusion, our results demonstrate that SAM-dPCR provides accurate, high-throughput, and reliable absolute quantification of biological samples. As the first self-supervised AI model applied in dPCR, SAM-dPCR eliminates the need for clean or 'ground truth' data, which is challenging due to the non-repetitive nature of each amplification process. This versatility positions SAM-dPCR as a powerful tool for researchers seeking accurate and efficient quantification of biological samples across different experimental setups. In the future, the SAM-dPCR pipeline can be adapted to multiplex tests in dPCR by designing fluorescent probes with different emission spectra specifically targeting different molecules or stratifying the fluorescence intensity for different targets. Moreover, optimizations could include further exploration into stain-less image analysis or virtual histological staining of biological samples for label-free and non-invasive diagnosis to further simplify the technical requirements of fluorescent dyes and the accompanying optics for detection.

## 4. Materials and methods
### 4.1 Sample preparation

High-fidelity Q5® DNA Polymerase (New England Biolabs) was employed as the primary commercial kit for the droplet digital PCR (ddPCR) reactions. The ddPCR reaction mix comprised diluted restriction enzymes, dNTPs, buffer, forward and reverse primers, Tween-20, PEG-8000, and PCR water. The detailed composition of the ddPCR reagents is outlined in **Supplementary Table 1**. The cDNA templates were serially diluted for the ddPCR experiments, maintaining a stock cDNA concentration of $(1.12 \pm 0.09) \times 10^4$ copies $\mu L^{-1}$.

For the ddPCR experiments depicted in **Fig. 2**, the PCR mixture was formulated with 1X KAPA HIFI buffer, 0.3 mM dNTP, 1X KAPA HIFI polymerase, 0.3 µM forward and reverse primers, templates, 0.1% NP-40, 0.2% Tween 20, and 0.1 mg/mL BSA (NEB, USA). Template concentrations of 40 pg, 4 pg, and 0.4 pg per 20 µL PCR system were used. The PCR protocol was initiated with denaturation at 95°C for 3 min, followed by 45 cycles of denaturation at 98°C for 20 s, annealing at 61°C for 15 s, and extension at 72°C for 15s. The final extension was conducted at 72°C for 1 min, with an indefinite hold at 12°C. The extracted Seahorse (Hippocampus kuda) genome, targeting the *cytochrome c oxidase subunit I (COI)* with an amplicon size of 206 bp, was utilized as the PCR template. The forward and reverse primers sequences were as follows:

Forward primer (5→3): TTTCTTCTCCTCCTTGCTTCCTCAG

Reverse primer (5→3): GAAATTGATGGGGGTTTTATGTTG

For the ddPCR experiments in **Fig. 3**, the PCR mixture incorporated 1X Platinum SuperFi II buffer, 0.2 mM dNTP, 1X Platinum SuperFi II polymerase, 0.5 µM forward and reverse primers, templates, 0.2% Tween 20, 0.2 mg/mL BSA (NEB, USA), and 0.4% PEG-8000. The PCR protocol included initial denaturation at 98°C for 30 s, followed by 25 or 45 cycles of denaturation at 98°C for 10 s, annealing at 60°C for 10 s, and extension at 72°C for 15s. The final extension was performed at 72°C for 5 min with an indefinite hold at 12°C. The sequences of the templates and primers used are detailed as follows:

The PCR template (5→3):

GTCTCGTGGAGCTCGACAGCATNNNNNNTGNNNNNNTGCCTACGACAAACAGACCTAAAATCGCTCATTGCATACTCTTCAATCAG

Forward primer (5→3):

Acrydite-ACTAACAATAAGCTCUAUAGTCTCGTGGAGCTCGACAG

Reverse primer (5→3):

CTGATTGAAGAGTATGCAATGAG

In the single-cell sequencing experiments (**Fig. 6**), the PCR mixture was prepared using 1X TaKaRa PrimeSTAR GXL buffer, 0.2 mM dNTP, 1X TaKaRa PrimeSTAR GXL polymerase, 0.2 µM forward and reverse primers, templates, 0.5% Tween 20, 0.1 mg/mL BSA (NEB, USA), and 0.5% PEG-8000. The PCR program was initiated with 25 cycles at 95°C for 10 s, 59°C for 15 s, and 68°C for 15 s, followed by a final extension at 68°C for 3 min and a hold at 25°C. The sequences of templates and primers are provided in Table 2.

**Table 2. Sequences of templates and primers in single-cell sequencing experiments**

|           |          |         | Sequence (5→3) |
|-----------|----------|---------|----------------|
| Barcode 1 | Template |         | GTCTCGTGGAGCTCGACAGNNNNNNNNNNNNTCGCTCATTGCATACTCTTCAATCAGC |
|           | Primer   | Forward | GTCTCGTGGAGCTCGACAG |
|           |          | Reverse | GCTGATTGAAGAGTATGCAATG |
| Barcode 2 | Template |         | GTCTCGTGAGTCAGGACAGNNNNNNNNNNNNTCGCTCATTGCATACTCTTCAATCAGC |
|           | Primer   | Forward | GTCTCGTGAGTCAGGACAG |
|           |          | Reverse | GCTGATTGAAGAGTATGCAATG |
| Barcode 3 | Template |         | GTCTCGTGACCTCGGACAGNNNNNNNNNNNNTCGCTCATTGCATACTCTTCAATCAGC |
|           | Primer   | Forward | GTCTCGTGACCTCGGACAG |
|           |          | Reverse | GCTGATTGAAGAGTATGCAATG |
| Barcode 4 | Template |         | GTCTCGTGGACAGTGACAGNNNNNNNNNNNNTCGCTCATTGCATACTCTTCAATCAGC |
|           | Primer   | Forward | GTCTCGTGGACAGTGACAG |

|  |  | Reverse | GCTGATTGAAGAGTATGCAATG |

### 4.2 Microfluidic chip fabrication and droplet generation

A flow-focusing microfluidic chip with cross-sectional dimensions of 30.0 μm (width) and 38.5 μm (height) was designed and fabricated for droplet generation 100 (**Fig. S3 Supplementary Information**). The chip was created using standard SU-8 photolithography and Polydimethylsiloxane (PDMS) replica molding processes. Initially, a master mold of 60.0 μm height was prepared by spinning SU8-2075 photoresist at 5000 r.p.m. for 28 s on a 4" silicon wafer, followed by UV exposure, baking, and developer bath according to the manufacturer's specifications. PDMS prepolymers (Dow, Inc., Sylgard 184) with a base-to-curing ratio of 10:1 were degassed, poured onto the SU-8 masters, and baked at 60°C for 10 h. The PDMS was then removed and cut to the desired shape, with inlets and outlets punched through. Both the PDMS slabs and glass slides were treated with oxygen plasma for 1 min and bonded by brief baking at 110°C. Finally, the microfluidic devices were hydrophobized by baking at 55°C for 24 h.

Monodisperse emulsions were generated using a custom microfluidic setup, driven by two syringe pumps (Legato 100, KD Scientific or Ph.D. 2000, Harvard Apparatus, USA) at the inlets. Commercial droplet generation oil (1864006, Bio-Rad Laboratories, Inc.) was used as the continuous phase. The resulting droplets exhibited a uniform size distribution with a mean diameter of 46.37 ± 1.64 µm (0.052 nL). The generated droplets were collected and transferred into 0.2 mL PCR stripe tubes, which were covered with mineral oil to minimize evaporation. PCR reactions were performed in a thermal cycler (Bio-Rad) with a hot start at 94 °C for 5 min, followed by 55 cycles of denaturation at 94 °C for 30 s, annealing, and extension at 60 °C for 1 min. Subsequently, the droplets were isolated onto a glass slide for observation.

### 4.3 Microreactors imaging and image processing

The amplified droplets were collected and dispensed into a specially designed PDMS chamber for observation under a lab fluorescence microscope. Imaging of the emulsions was performed using an inverted microscope (Eclipse Ti-U, Nikon) coupled with a camera (DS-Qi2, Nikon) in both brightfield and fluorescence fields. Fluorescence excitation was achieved at 455 nm, and the emitted light was captured by a CCD through a 495 nm long-pass filter. Images were acquired at 4× magnification with an exposure time of 1 s and a sensor sensitivity of 200. For each sample, 10 images were taken at different positions, excluding edges, encompassing over 2,000 droplets in total. The droplet diameter was quantified and analyzed using ImageJ (National Institutes of Health and the Laboratory for Optical and Computational Instrumentation).

The microwell droplet digital PCR (dPCR) images presented in this manuscript were acquired using a 3D Digital PCR chip v2, with each reaction well having a volume of 755 pL (ThermoFisher Scientific, USA). These experiments were conducted by our collaborator, Prof. Mingli You, from the School of Life Science and Technology at Xi'an Jiaotong University[31]. The experimental design involved a series of dilutions, spanning a broad dynamic range of template concentrations, i.e., negative, 100×, 20×, 10× and 5× dilutions from the original concentration of $1.66 \times 10^{-13}$ mol/L. This study amplified two types of double-stranded genes, specifically blaNDM and blaVIM, which regulate the expression of β-lactamases, a class of antibiotic agents used against carbapenems. The corresponding DNA sequences can be accessed on the GenBank website (https://www.ncbi.nlm.nih.gov/genbank/) using the nucleotide codes NC023908 and NC023274 for blaNDM and blaVIM, respectively. Due to the extensive length of the complete gene sequences, we selected specific subsequences as templates and subsequently redesigned the corresponding primers and probes. These genes were synthesized as plasmids by QingsKe, based in Xi'an, China, who also synthesized all primers. The probes for blaNDM and blaVIM, labeled as blaNDM-AF488 and blaVIM-TET respectively, were modified with AF488 and TET, and were synthesized by Sangon, based in Shanghai, China.

The original agarose digital PCR images depicted in **Fig. 6 (c)** were provided by our collaborators, Dr. Xuefei Leng and Prof. Chaoyong Yang, from the College of Chemistry and Chemical Engineering at Xiamen University[35]. The original images in **Fig. 6 (d)** were extracted from previously published bacterial quantification work[36].

### 4.4 Deep-learning-assisted automatic data analysis

The SAM-dPCR algorithm utilized the Zero-Shot Segment Anything Model (SAM) as its core architecture. SAM is a self-supervised deep learning model that enables accurate image segmentation without the need for annotated training data. In this study, we employed the ViT-B encoder variant of SAM due to its balance between model size and performance. The ViT-B encoder has 91 million parameters, providing a good trade-off between computational efficiency and segmentation accuracy. The entire process was executed using the Pytorch deep learning framework on an NVIDIA Tesla V100-SXM2-16GB hardware platform.

This self-supervised SAM-dPCR deep learning model segmented and classified microreactors, negating manual annotation. Each of the segmented reactors was evaluated based on diameter, predicted Intersection over Union (IoU), and stability score simultaneously. Statistical analysis was then performed, including distribution and inferred concentration based on fluorescence intensity differences. The figure displays images from both droplet and microwell dPCR benchtop experiments, overlaid with segmentation masks that are automatically annotated by the zero-shot SAM model. These masks delineate the microreactors and classify them into positive and negative categories based on fluorescence intensity.

The performance of ddPCR was assessed by evaluating its linearity range, limit of quantification (LOQ), and reproducibility. Sample concentration inference was determined by fitting the data into a Poisson distribution using Equation (8). Over 2,000 microreactors, including droplets and microwells, were analyzed for each concentration.

The probability $\Pr(X = k)$ that a microreactor will contain k copies of target gene if the mean number of target copies per microreactor is $\lambda$:

$$f(k, \lambda) = \Pr(X = k) = \frac{\lambda^k e^{-\lambda}}{k!} \tag{1}$$

where

$k$ is the number of occurrences ($k$ can take values 0, 1, 2, ...).

$e$ is Euler's number ($e = 2.71828…$).

! is the factorial function.

Inputting $k=0$ gives the probability that a microreactor will be empty:

$$\Pr(X = 0) = e^{-\lambda} \tag{2}$$

For the number of microreactors being large enough, the observed fraction of empty microreactors (E) gives estimation of $\Pr(X = 0)$

$$E = e^{-\lambda} \tag{3}$$

At the same time, by definition of E,

$$E = \frac{N_{negative}}{N} \tag{4}$$

Solving (3) we get

$$\lambda = -\ln(E) \tag{5}$$

As $\lambda$ is the copies per microreactor, concentration of copies per volume is

$$Concentration = \frac{\lambda}{V_{microreactor}} \tag{6}$$

Which means

$$Concentration = \frac{-\ln(E)}{V_{microreactor}} \tag{7}$$

Combining (4) and (7), we get

$$Concentration = -\ln\left(\frac{N_{negative}}{N_{total}}\right)/V_{microreactor} \tag{8}$$

### 4.5 Graphical user interface

The SAM-dPCR algorithm was implemented as a standalone software tool using Python and packaged as a user-friendly graphical user interface (GUI). The GUI allowed users to interact with the software, input dPCR images, and visualize the segmentation results in real-time. The software was designed to seamlessly integrate with common laboratory fluorescence microscopes, enabling easy adoption and utilization in different experimental setups.

The GUI also plots the droplets by accumulating results from sequential frames. Users have the option to digitally save the raw images, background-subtracted images, plot results, and calculated values. The size of the droplets and the calculated template concentration are continuously displayed at the bottom. Additionally, the GUI can operate in offline mode to analyze pre-saved image datasets by reading folders. The design codes for the GUI can be found in the **Supporting Materials**.

### 4.6 Statistical analysis

Statistical analysis was performed using GraphPad Prism software (GraphPad Software). All data are presented as mean ± standard deviation (SD) with n ≥ 3. Hypothesis testing was conducted using a t-test, and significance was defined as $p \leq 0.05$.

## 5. References

1. Jin, M. *et al.* StratoLAMP: Label-free, multiplex digital loop-mediated isothermal amplification based on visual stratification of precipitate. *Proceedings of the National Academy of Sciences* **121**, (2024).

2. Yin, H. *et al.* Ultrafast multiplexed detection of SARS-CoV-2 RNA using a rapid droplet digital PCR system. *Elsevier*.

3. Zhang, W. *et al.* Pipette-Tip-Enabled Digital Nucleic Acid Analyzer for COVID-19 Testing with Isothermal Amplification. *Anal Chem* **93**, 15288–15294 (2021).

4. Hindson, C. M. *et al.* Absolute quantification by droplet digital PCR versus analog real-time PCR. *Nat Methods* **10**, 1003–1005 (2013).

5. Low, H., Chan, S. J., Soo, G. H., Ling, B. & Tan, E. L. Clarity$^{TM}$ digital PCR system: a novel platform for absolute quantification of nucleic acids. *Anal Bioanal Chem* **409**, 1869–1875 (2017).

6. Mock, U., Hauber, I. & Fehse, B. Digital PCR to assess gene-editing frequencies (GEF-dPCR) mediated by designer nucleases. *Nature Protocols 2016 11:3* **11**, 598–615 (2016).

7. Veyer, D. *et al.* HPV circulating tumoral DNA quantification by droplet-based digital PCR: A promising predictive and prognostic biomarker for HPV-associated oropharyngeal cancers. *Int J Cancer* **147**, 1222–1227 (2020).

8. Strain, M. C. *et al.* Highly Precise Measurement of HIV DNA by Droplet Digital PCR. *PLoS One* **8**, e55943 (2013).

9. Geng, Z. *et al.* "Sample-to-Answer" Detection of Rare ctDNA Mutation from 2 mL Plasma with a Fully Integrated DNA Extraction and Digital Droplet PCR Microdevice for Liquid Biopsy. *Anal Chem* **92**, 7240–7248 (2020).

10. Postel, M., Roosen, A., Laurent-Puig, P., Taly, V. & Wang-Renault, S. F. Droplet-based digital PCR and next generation sequencing for monitoring circulating tumor DNA: a cancer diagnostic perspective. *Expert Rev Mol Diagn* **18**, 7–17 (2018).

11. Quan, P. L., Sauzade, M. & Brouzes, E. DPCR: A technology review. *Sensors (Switzerland)* vol. 18 Preprint at https://doi.org/10.3390/s18041271 (2018).

12. Wei, Y., Cheng, G., Ho, H. P., Ho, Y. P. & Yong, K. T. Thermodynamic perspectives on liquid-liquid droplet reactors for biochemical applications. *Chem Soc Rev* **49**, 6555–6567 (2020).

13. Chen, Z. *et al.* Centrifugal micro-channel array droplet generation for highly parallel digital PCR. *Lab Chip* **17**, 235–240 (2017).

14. Zhu, Q. *et al.* Digital PCR on an integrated self-priming compartmentalization chip. *Lab Chip* **14**, 1176–1185 (2014).

15. Baker, M. Digital PCR hits its stride. *Nat Methods* **9**, 541–544 (2012).

16. Wu, Z. *et al.* Absolute quantification of DNA methylation using microfluidic chip-based digital PCR. *Biosens Bioelectron* **96**, 339–344 (2017).

17. Zhang, W. *et al.* An Integrated ddPCR Lab-on-a-Disc Device for Rapid Screening of Infectious Diseases. *Biosensors 2024, Vol. 14, Page 2* **14**, 2 (2023).

18. Pinheiro, L. B. *et al.* Evaluation of a droplet digital polymerase chain reaction format for DNA copy number quantification. *Anal Chem* **84**, 1003–1011 (2012).

19. Chen, B. *et al.* Droplet digital PCR as an emerging tool in detecting pathogens nucleic acids in infectious diseases. *Clinica Chimica Acta* **517**, 156–161 (2021).

20. Chen, J. *et al.* Capillary-based integrated digital PCR in picoliter droplets. *Lab Chip* **18**, 412–421 (2018).

21. Bartkova, S., Vendelin, M., Sanka, I., Pata, P. & Scheler, O. Droplet image analysis with user-friendly freeware CellProfiler. *Analytical Methods* **12**, 2287–2294 (2020).

22. Bai, B. *et al.* Deep learning-enabled virtual histological staining of biological samples. *Light: Science and Applications* vol. 12 Preprint at https://doi.org/10.1038/s41377-023-01104-7 (2023).

23. Wei, Y. *et al. Deep-DGFP: Deep Learning Approach for Large-Scale, Real-Time Quantification of Green Fluorescent Protein-Labeled Biological Samples in Microreactors*.

24. Li, Y. *et al.* Incorporating the image formation process into deep learning improves network performance. *Nat Methods* **19**, 1427–1437 (2022).

25. Bai, B. *et al.* Label-Free Virtual HER2 Immunohistochemical Staining of Breast Tissue using Deep Learning. *BME Front* **2022**, (2022).

26. Gardner, K. *et al.* Deep learning detector for high precision monitoring of cell encapsulation statistics in microfluidic droplets. *Lab Chip* **22**, 4067–4080 (2022).

27. Nawaz, M. *et al.* Unraveling the complexity of Optical Coherence Tomography image segmentation using machine and deep learning techniques: A review. *Computerized Medical Imaging and Graphics* vol. 108 Preprint at https://doi.org/10.1016/j.compmedimag.2023.102269 (2023).

28. Hu, Z. *et al.* A novel method based on a Mask R-CNN model for processing dPCR images. *Analytical Methods* **11**, 3410–3418 (2019).

29. Wei, Y. *et al.* Deep-qGFP: A Generalist Deep Learning Assisted Pipeline for Accurate Quantification of Green Fluorescent Protein Labeled Biological Samples in Microreactors. *Small Methods* (2023) doi:10.1002/smtd.202301293.

30. Yang, H. *et al.* A deep learning based method for automatic analysis of high-throughput droplet digital PCR images. *Analyst* **148**, 239–247 (2022).

31. Cao, C. *et al.* Similar color analysis based on deep learning (SCAD) for multiplex digital PCR via a single fluorescent channel. *Lab Chip* **22**, 3837–3847 (2022).

32. Anagnostidis, V. *et al.* Deep learning guided image-based droplet sorting for on-demand selection and analysis of single cells and 3D cell cultures. *Lab Chip* **20**, 889–900 (2020).

33. Chen, L., Ding, J., Yuan, H., Chen, C. & Li, Z. Deep-dLAMP: Deep Learning-Enabled Polydisperse Emulsion-Based Digital Loop-Mediated Isothermal Amplification. *Advanced Science* **9**, 1–9 (2022).

34. Kirillov, A. *et al.* Segment Anything. (2023).

35. Leng, X., Zhang, W., Wang, C., Cui, L. & Yang, C. J. Agarose droplet microfluidics for highly parallel and efficient single molecule emulsion PCR. *Lab Chip* **10**, 2841–2843 (2010).

36. Geersens, É., Vuilleumier, S. & Ryckelynck, M. Growth-Associated Droplet Shrinkage for Bacterial Quantification, Growth Monitoring, and Separation by Ultrahigh-Throughput Microfluidics. *ACS Omega* **7**, 12039–12047 (2022).

## 6. Acknowledgements

The authors are grateful to the funding support from the Hong Kong Research Grants Council (project reference: GRF14204621, GRF14207121, GRF14207920, GRF14207419, GRF14203919, N_CUHK407/16), the Marine Conservation Enhancement Fund (MCEF20108_L02), and the Innovation and Technology Commission (project reference: GHX-004-18SZ).

The authors would like to acknowledge Prof. Mingli You (School of Life Science and Technology, Xi'an Jiaotong University), Dr. Xuefei Leng and Prof. Chaoyong Yang (College of Chemistry and Chemical Engineering, Xiamen University), Dr. Ronjie Zhao and Dr. Meng Yan (State Key Laboratory of Marine Pollution, City University of Hong Kong, Hong Kong, China), Mr. Syed Muhammad Tariq Abbasi, Mr. Minqing Zhang, Dr. Shiyue Liu, Dr. Guangyao Cheng, Dr. Shutian Zhao, Dr. Md Habibur Rahman, Mr. Chenglang Yuan, Mr. Shirui Zhao, Ms. Khadija BIBI, and Ms. Syeda Aimen Abbasi (Department of Biomedical Engineering, The Chinese University of Hong Kong) for their support in the project development.

## 7. Author information

Authors and Affiliations

**Department of Biomedical Engineering, The Chinese University of Hong Kong, Shatin, Hong Kong SAR, 999077, China**

Yuanyuan Wei, Fuyang Qu, Yingqi Fu, Guangyao Cheng, Yi-Ping Ho, Ho-Pui Ho & Wu Yuan

**Department of Biomedical Engineering, National University of Singapore, 119077, Singapore**

Shanhang Luo

**Department of Computer Science and Engineering, The Chinese University of Hong Kong, Shatin, Hong Kong SAR, 999077, China**

Changran Xu

**Department of Electronic Engineering, The Chinese University of Hong Kong, Shatin, Hong Kong SAR, 999077, China**

Yi Zhang

**Centre for Biomaterials, The Chinese University of Hong Kong, Hong Kong SAR, 999077, China**

Yi-Ping Ho

**Hong Kong Branch of CAS Center for Excellence in Animal Evolution and Genetics, Hong Kong SAR, 999077, China**

Yi-Ping Ho

**State Key Laboratory of Marine Pollution, City University of Hong Kong, Hong Kong SAR, 999077, China**

Yi-Ping Ho

Contributions

Y. Wei, and S. Luo and Y. Wu contributed to the study's conception and design, with S. Luo specifically realized the software functions and implementation. Y. Wei and F. Qu conducted the biological experiments. Y. Wei, S. Luo, C. Xu, and Y. Fu performed the image analysis experiments, data analysis and figures editing. F. Qu, G. Cheng, and Y. Ho supplied the microfluidic chip and customized microfluidic platform. Y. Wei wrote the manuscript with contributions from all authors. H. Ho and W. Yuan conceived the project and supervised the research.

Corresponding author

Correspondence to Ho-Pui Ho and Wu Yuan.

## 8. Ethics declarations

No conflict of interest

## 9. Additional information

Electronic supplementary material

Supplementary Information